\documentclass{aastex}
\usepackage{emulateapj5}

\def\'#1{\ifx#1i{\accent"13\i}\else{\accent"13#1}\fi}
\def\VS{V\'azquez-Semadeni}

\newif\ifAMStwofonts
\begin{document}

\title{The Probability Distribution Function of Column Density in
Molecular Clouds}
[The PDF of Column Density in Molecular Clouds]

\author{Enrique V\'azquez-Semadeni$^1$ and Nieves Garc\'ia$^2$}
      [V\'azquez-Semadeni \& Garc\'ia]

\affil{$^1$Instituto de Astronom\'\i a, UNAM, Campus Morelia,
Apdo. Postal 3-72, Xangari, 58089, Morelia, Mich., MEXICO} 
\affil{$^2$Instituto de Astronom\'\i a, UNAM, Apdo. Postal 70-264,
M\'exico D.F., 04510, MEXICO}


%

\begin{abstract}
We discuss the probability distribution function (PDF) of column
density resulting from density fields with lognormal PDFs, applicable
to isothermal gas (e.g., probably molecular clouds). For magnetic and
non-magnetic numerical simulations 
of compressible, isothermal turbulence forced at intermediate scales
(1/4 of the box size), we find that the autocorrelation 
function (ACF) of the density field decays over relatively short
distances compared to the simulation size. We suggest that a
``decorrelation length'' can be defined as the distance over which the
density ACF has decayed to, for example, 10\% of its zero-lag value, so
that the density ``events'' along a line of sight can be assumed 
to be independent over distances larger than this, and the
Central Limit Theorem should be applicable. However, using random
realizations of lognormal fields, we show that 
the convergence to a Gaussian is extremely slow in the high-density tail. 
As a consequence, the column density PDF is not expected to
exhibit a unique functional shape, but to transit
instead from a lognormal to a Gaussian form as the ratio $\eta$ of the 
column length to the decorrelation length (i.e., the number of
independent events in the cloud) increases. Simultaneously, the PDF's
variance decreases. 
For intermediate values of $\eta$, the column density PDF assumes  
a nearly exponential decay. For cases with a density
contrast of $10^4$ (resp.\ $10^6$), as found in intermediate-resolution
simulations, and expected from GMCs to dense molecular cores,
the required value of $\eta$ for convergence to a Gaussian is at least a
few hundred (resp.\ 
several thousand). We then discuss the density power spectrum and the
expected value of $\eta$ in actual molecular clouds, concluding that they 
are uncertain since they may depend on several physical parameters.

Observationally, our results suggest that $\eta$ may be inferred from the
shape and width of the column density PDF in optically-thin-line or
extinction studies. Our results should also hold for gas with
finite-extent power-law underlying density PDFs, which should be
characteristic of the diffuse, non-isothermal neutral medium (temperatures
ranging from a few hundred to a few thousand degrees). Finally, we note
that for $\eta \gtrsim 100$, the dynamic range in column density is small
($\lesssim$ a factor of 10), but this is only an averaging effect, with
no implication on the dynamic range of the underlying density
distribution.

\end{abstract}


\section{Introduction}

In recent years, several studies of the probability density
function\footnote{The PDF is frequently also referred to, in loose form,
as the probability distribution function. Note also that the PDF is a
{\it one-point} statistic and 
contains no spatial information, contrary to the case of, say, the
correlation function, which is a {\it two-point} statistic, and from which
the PDF is an independent quantity.} (PDF) of the density field in
numerical simulations of compressible turbulent flows have been 
advanced as a first step in its full statistical
characterization. These studies have shown that the density PDF
depends on the effective polytropic exponent $\gamma$ of
the fluid, defined by the expression $P \propto \rho^\gamma$, where
$P$ is the
pressure and $\rho$ is the gas density. Specifically, for isothermal flows
($\gamma=1$), the PDF is lognormal (V\'azquez-Semadeni 1994; Padoan,
Nordlund \& Jones 1997; Passot \& V\'azquez-Semadeni 1998; Scalo et
al.\ 1998; Ostriker, Gammie \& Stone 1999; Ostriker,
Stone \& Gammie 2000), while Passot \& V\'azquez-Semadeni (1998) noted
that, for $\gamma <1$ (resp.\ $\gamma>1$), the 
PDF develops a power-law tail at high (resp.\ low) densities (see also
Scalo et al.\ 1998,
Nordlund \& Padoan 1999, and the review by V\'azquez-Semadeni et
al.\ 2000). Additionally, Gotoh \& Kraichnan (1993) have reported a
power-law tail at high densities for Burgers flows, and Porter,
Pouquet \& Woodward (1991) have reported an exponential behavior for
adiabatic flows. Passot \& V\'azquez-Semadeni (1998) explained
the lognormal PDF for isothermal flows as a consequence of
the Central Limit Theorem (CLT) acting on the distribution of the logarithm
of the density field. They assumed that a given density distribution is
arrived at by a succession of multiplicative density jumps, which are
therefore additive in the logarithm. Since for an isothermal flow the
speed of sound is spatially uniform, the density jump expected
from a shock of a given strength is independent of the local density,
and thus all density jumps can be assumed to follow the same
distribution (determined by the distribution of Mach
numbers, as studied, for example, by Smith, Mac Low \& Zuev 2000 and
Smith, Mac Low 
\& Heitsch 2000). Finally, at a given position in space, each density jump is
independent of the previous and following ones. Therefore, the CLT,
according to which the distribution of the sum of
identically-distributed, independent events approaches a Gaussian,
can be applied to the logarithm of the density, and the density itself 
is expected to possess a lognormal PDF.

However, the observationally accessible quantity is not the PDF of the 
mass (or ``volume'') density, but rather that of the {\it column
density}, i.e., the integral (or sum, for a discrete spatial grid) of the
density along one spatial dimension (the ``line of sight'', or LOS).
Recently, Padoan et al.\ (2000) and Ostriker et al.\ (2000, hereafter
OSG01) have also discussed this PDF in 
three-dimensional (3D) numerical simulations of isothermal
compressible MHD turbulence with resolutions up to $256^3$ zones. In
particular, OSG01 have
found that the column density distribution has essentially the same
shape as that of the underlying density field (a lognormal for
isothermal gas), although with smaller mean and width. This
result is puzzling because, according to the CLT, the PDF of column
density should approach a Gaussian shape, the column density being
proportional to the mean density along the LOS. OSG01 attributed the
apparent inapplicability of the CLT to the possible presence of
intermediate-sized structures in the density
field that invalidate the requirement of statistical independence
of the individual zones needed for the CLT. 

In this paper we suggest that density ``events'' along the LOS can be
regarded as independent if they are separated by distances larger than
some ``decorrelation length'', over which the density autocorrelation
function (ACF) has decayed by a large enough factor (we use a factor
of 10). If the column length is significantly larger than this, then 
convergence to a Gaussian might be expected. For 3D numerical
simulations of magnetic and 
non-magnetic, isothermal turbulence forced at intermediate-to-large
scales (1/4 of the box size), we find that the ACF drops to the 10\% 
level at relatively short separations ($\sim 15$\% of the box size).
Using random 
realizations of 3D lognormal fields, we show that this convergence is
nevertheless very slow because of the large skewness (asymmetry) and
kurtosis (wing excess) of the lognormal density PDF. Then we discuss in
a speculative way the factors that may determine the shape of the
density ACF in molecular clouds, and suggest that its
characteristic length may be inferred observationally.
In \S \ref{sec:num_dat} we describe the numerical data we use, both
from simulations of isothermal compressible turbulence and from random
realizations of lognormal fields. In \S \ref{sec:pdfs} we discuss the
ACFs and PDFs of the projected density fields and, in particular, the
LOS lengths required for convergence to a Gaussian. In \S
\ref{sec:discussion} we discuss the PDF width in simulations and
observations, the case of non-isothermal gas, the dependence of the
correlation length on physical parameters of the turbulence, and some
caveats. Finally, in \S \ref{sec:summary} we summarize our results.

\section{Numerical data} \label{sec:num_dat}

We use two different sets of data for our analysis. The first
comprises two numerical simulations of forced, compressible,
isothermal, 3D
turbulence, performed at a resolution of $100^3$ grid points, one
non-magnetic and one magnetic. The
numerical method is pseudospectral with periodic boundary conditions,
employing a combination of
eighth-order hyperviscosity and second-order viscosity which allows
larger turbulent inertial ranges than can be attained with
second-order viscosity only. A second-order mass diffusion operator is 
included as well. We refer the reader to Passot, V\'azquez-Semadeni \& 
Pouquet (1995) and V\'azquez-Semadeni, Passot \& 
Pouquet (1996) for details. Here we just mention that for both runs
the forcing rises as $k^4$ for $2 \le k \le 4$, and decays as $k^{-4}$
for $4 < k \le 15$, where $k$ is the wavenumber. For the non-magnetic 
run the forcing is 100\% compressible and has an amplitude of 25 in
code units; the
hyperviscosity coefficient $\nu$ is $8 \times 10^{-11}$, the
second-order coefficient $\mu$ is $3.56 \times 10^{-3}$, and the mass
diffusion coefficient $\mu_\rho$ is $0.02$. The sound speed is $c=0.5
u_0$, where $u_0$ is the velocity unit. The Mach number has an rms 
value $\sim 1$, with maximum excursions up to $\sim 3.5$. For the
magnetic run the forcing also peaks at $k=4$, but is 50\%  
compressible, and has an amplitude of 7.5 in code units; the
diffusive coefficients are $\nu=2 \times 10^{-11}$, $\mu =
3.5 \times 10^{-3}$, and $\mu_\rho=0.03$. The sound speed is $c=0.2
u_0$, giving an rms Mach number $\sim 2.5$. A uniform magnetic 
field is placed initially along the $x$
direction, giving a $\beta$ parameter,
defined as the ratio of the mean thermal to magnetic pressures, equal
to 0.04, and an rms Alfv\'enic Mach number $\sim 0.5$. We 
have chosen this rather strongly magnetized case in order to bring out 
the effects of the magnetic field clearly. The differences between the
two simulations are due to the fact that
the magnetic simulation was not originally intended for the present
study, but we do not believe this is a concern for our purposes.
Our simulations are only mildly supersonic because of limitations of
both the numerical scheme and the computational resources available to
us, which constrain the resolution to the value mentioned above. 

Since at $100^3$ a projection along one axis gives a square of only
$100^2$ points, column density PDFs for one single temporal snapshot
contain only 10,000 data points, giving relatively poor
statistics. We thus take advantage of the fact that the simulations are
statistically stationary (although the maximum density contrast and rms Mach
number do fluctuate by about 50\% in time), and choose
to combine several density snapshots to produce a single column density 
histogram. Specifically, for the non-magnetic run we use 19 subsequent 
snapshots, spaced an amount $\Delta t=0.1$ code time units ($\sim 1.6
\times 10^{-2}$ large-scale turbulent crossing times at the rms speed). For the
magnetic run we use 18 snapshots, spaced an amount $\Delta t=0.2$ code
units ($3.2 \times 10^{-2}$ large-scale turbulent crossing times).  

In order to overcome the limitations of the numerical simulations, we
consider a second set of data, consisting of simple realizations of
random fields with 
lognormal PDFs, obtained by generating random numbers $X_i$ with a standard
Gaussian distribution (zero mean and unit variance) and defining
a new random variable $\rho_i=e^{b X_i}$, where $b$ is a parameter that
controls the width of the lognormal distribution. We use sequences of these
``density'' values to fill ``cubes'' (actually parallelepipeds) with
fixed ``plane of the sky'' (POS) dimensions $\Delta x$ and $\Delta y$, and
``LOS'' lengths $\Delta z$ ranging from a few tens of grid cells to a few
thousands. 

It is important to note that we have two different sets of 
``samples'' in this problem: one is the set of points along the
LOS, whose number is given by $\Delta z$ (for simplicity, $\Delta z$
is measured in grid
cells, so that it is numerically equal to the number of contributing
cells). The density is effectively{\it  averaged} along the LOS. The other
sample is the set of lines of sight in the POS, whose number is
given by the product $\Delta x \Delta y$. This equals the
number of data points in the column
density PDFs. We emphasize that the number of points in a PDF is
completely independent of the LOS length $\Delta z$, so that we
can have PDFs with the same number of data points, but with different
values of $\Delta z$. Increasing the number of points in the POS
allows us to improve the ``signal-to-noise'' ratio for the
PDF, especially at the wings. However, the functional {\it shape}
of the PDF is expected to 
depend only on the number of points in the LOS. Indeed, the column
density is equivalent to the {\it sample mean} (along the LOS) in
sampling theory, and it is well known that the statistics of the sample mean
depend on the sample size (again, the sample along the LOS). In other
words, the column density PDFs are histograms of the sample
means, of which there is one for each LOS.

To improve the PDF signal-to-noise ratio, we consider many 
parallelepipeds (actually, 50 in all cases, each with $\Delta x=\Delta 
y=50$) for each set of parameters ($b,\Delta z$), although 
this is exactly equivalent to having a single larger parallelepiped with
125,000 data points in the ``plane of the sky'', due to the statistical
independence of the data, and we only keep track of the individual
parallelepipeds for analogy with the procedure of combining several
temporal snapshots used in the case of the numerical simulations. But
in practice, the only relevant datum in this
sense is how many data points does each PDF contain, the projected ``shape''
of the parallelepiped on the POS being completely irrelevant (for
example, it may be a square, or a straight line). Thus, the total
number of grid cells in the larger parallelepipeds (i.e., their total
volume), is 125,000$\times\Delta z$.
We consider two subsets of data, obtained from
using two different values of $b$, namely $b=1$ and
$b=1.5$. 

For both the simulation and the random data sets, we first normalize
the lognormal density data as required by the CLT, by defining a new
variable $\rho_i^\prime \equiv (\rho_i-\langle \rho
\rangle)/\sigma_\rho$, where 
$\langle \rho \rangle$ is the mean density, $\sigma_\rho$ is the
standard deviation, and $i$ counts pixel position along the LOS. For
the random data, the mean and variance of the $\rho$
distribution are related to those of the Gaussian variable $X$ by
$\langle \rho \rangle = \exp(\langle X \rangle + \sigma_X^2/2)$ and
$\sigma_\rho^2 = \bigl[\exp(\sigma_X^2)-1\bigr] \exp(2\langle X\rangle +
\sigma_X^2)$ (see, e.g., Peebles 1987, app.\ F). For the simulation data, the
mean density is 1, but $\sigma_\rho$ is not known {\it a priori}, and
attempting to measure it gives large errors both because of the
relatively high frequency of high-density events and because it is not 
constant over time. We find
empirically that the necessary values of $\sigma_\rho$ to bring the
column density to near unit variance (see below) are approximately 2
and 3 for the non-magnetic and magnetic runs, respectively.

\begin{figure}
\plottwo{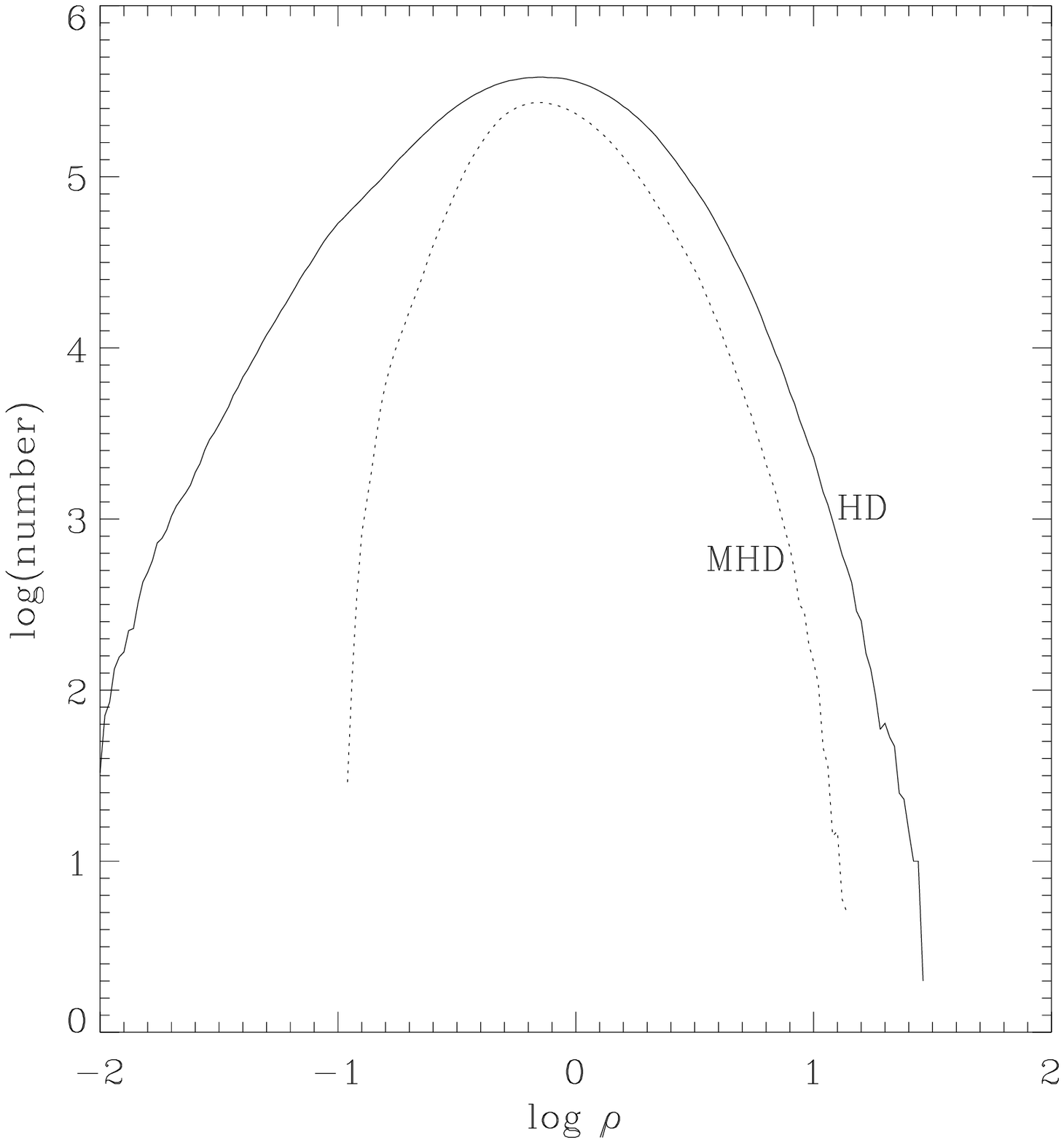}{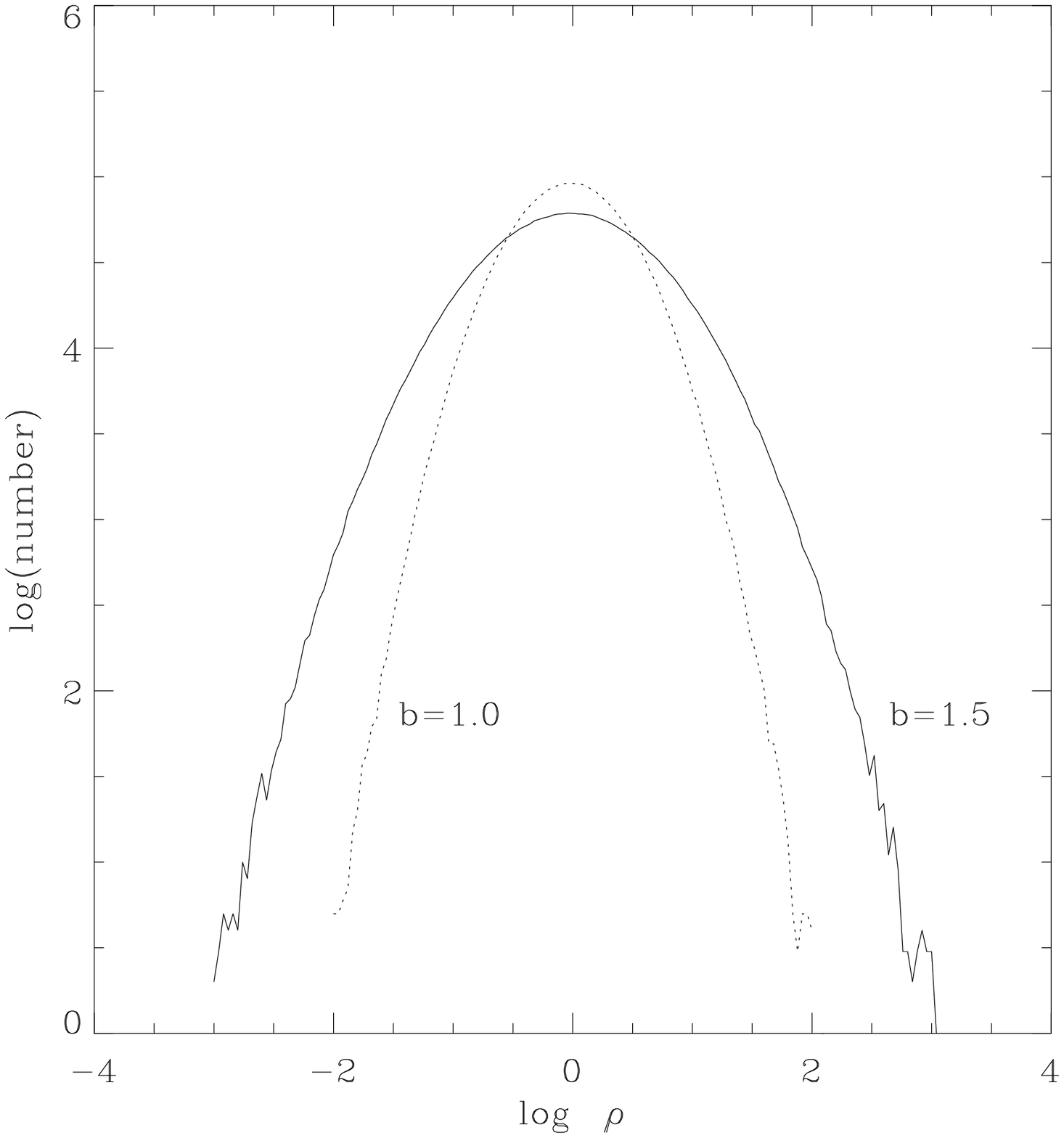}
\caption{a) ({\it Left}) Density PDFs of the non-magnetic ({\it solid
line}) and magnetic ({\it dotted line}) simulations. b) ({\it Right})
Density PDFs of the random realizations, for $b=1.5$ ({\it solid
line}) and $b=1$ ({\it dotted line}).}
\label{fig:rho_pdfs}
\end{figure}

We then project (sum) the normalized density along the $z$-axis to
obtain its associated normalized (i.e., of zero-mean and
unit-variance) ``column density'' $\zeta$, defined by (Peebles 1987,
sec.\ 4.7)
\begin{equation}
\zeta \equiv \frac{\sum_i \rho_i^\prime}{\sqrt{\Delta z}},
\label{eq:col_dens}
\end{equation}
where the sum extends over all grid cells along the LOS. In the next section
we discuss the PDFs of $\zeta$. Figure \ref{fig:rho_pdfs} shows the
underlying density PDFs for the numerical simulation data ({\it left})
and for the random lognormal data ({\it right}) before
normalization. The density fields are seen to be
exactly lognormal in the case of the random data, and approximately so
in the simulation data. The PDF of the non-magnetic simulation
exhibits an excess at
small densities but, since we will be focusing
mostly on the high-density side, and our main conclusions will be
drawn from the random data, we do not consider this excess to be a
problem. Note also that the non-magnetic run has a wider density PDF
even though it has a smaller mean Mach number than its magnetic
counterpart. This is probably due to the fact that in the latter the
forcing is only 50\% compressible, and of smaller amplitude. The
density PDFs for the random data are seen to span dynamic 
ranges of $10^4$ and $10^6$ for $b=1$ and $b=1.5$ respectively.

\section{The ACFs and column density PDFs} \label{sec:pdfs}

\begin{figure}
\plotone{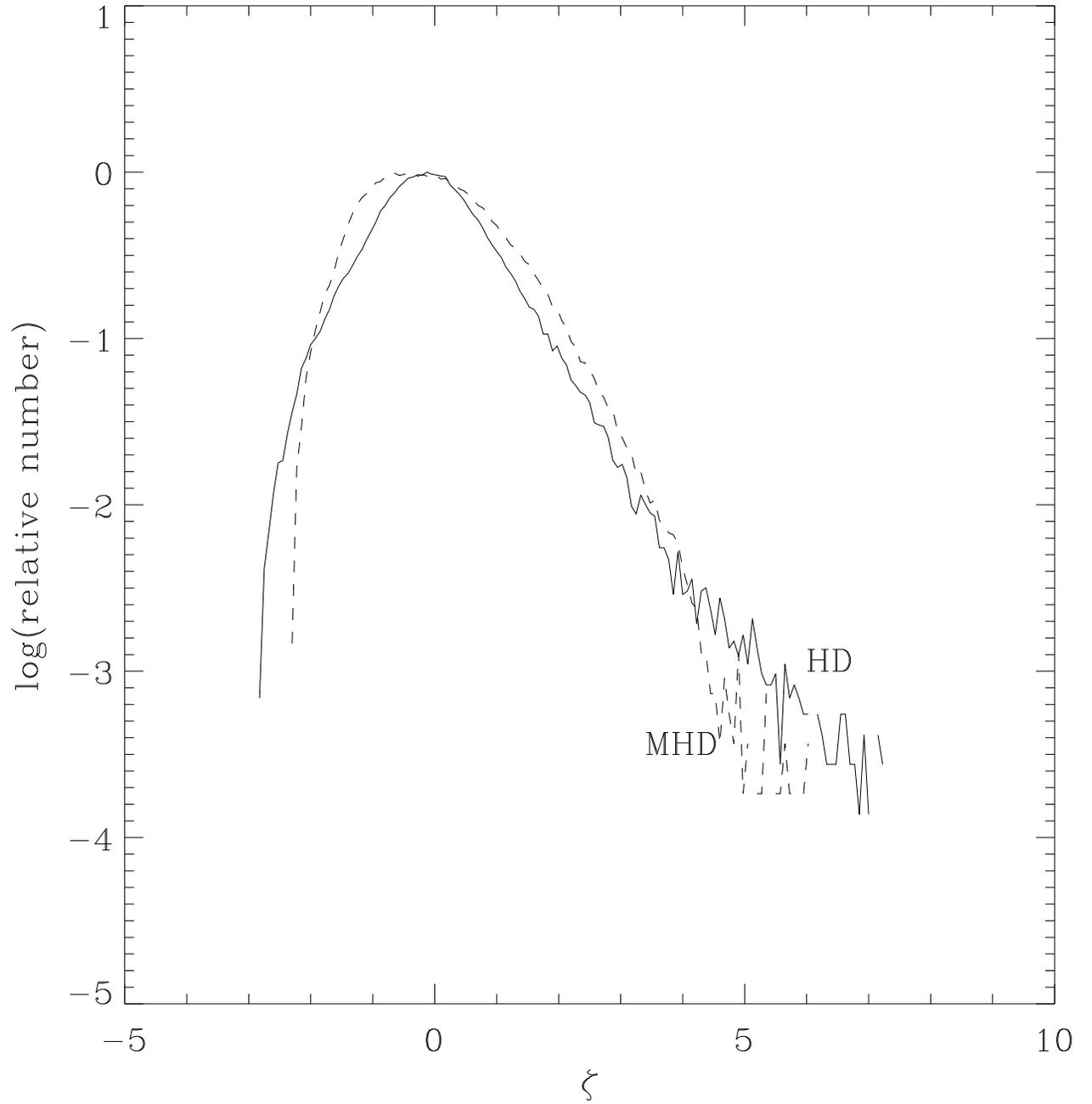}
\caption{PDFs of {\it normalized} column density $\zeta$ obtained from
all lines of sight and combining several snapshots as indicated in the 
text. ({\it Solid line}):  non-magnetic run. ({\it Dotted line}):
magnetic run.}
\label{fig:N_pdfs_num}
\end{figure}

Figure \ref{fig:N_pdfs_num} shows, in log-$y$, lin-$x$ form, the
time-integrated (i.e., adding 
several temporal snapshots into the same histogram) normalized-column
density ($\zeta$) PDFs for the magnetic and non-magnetic numerical
simulations. In this graph format, a Gaussian is a parabola, and an
exponential is a straight line. In the two runs, a
nearly exponential decay is apparent at moderately high $\zeta$, although
the very-large-$\zeta$ tail
clearly exhibits an excess from this trend in the non-magnetic case
and a defect in the magnetic one. This may be an effect of the less
extended underlying density PDF in the magnetic case. As already
pointed out by OSG01 with respect to their nearly lognormal column
density PDFs, these results are puzzling: one
would expect the $\zeta$-PDF to be Gaussian, as the column density is
essentially a sum (or equivalently, an average) of the density events
along each LOS, whose
distribution should approach a Gaussian by virtue of the CLT. As
mentioned in the Introduction, OSG01 interpreted the deviation from
Gaussianity in terms of a violation of the statistical independence
requirement of the CLT, due to the existence of intermediate-size
correlations in the density field.

\begin{figure}
\plotone{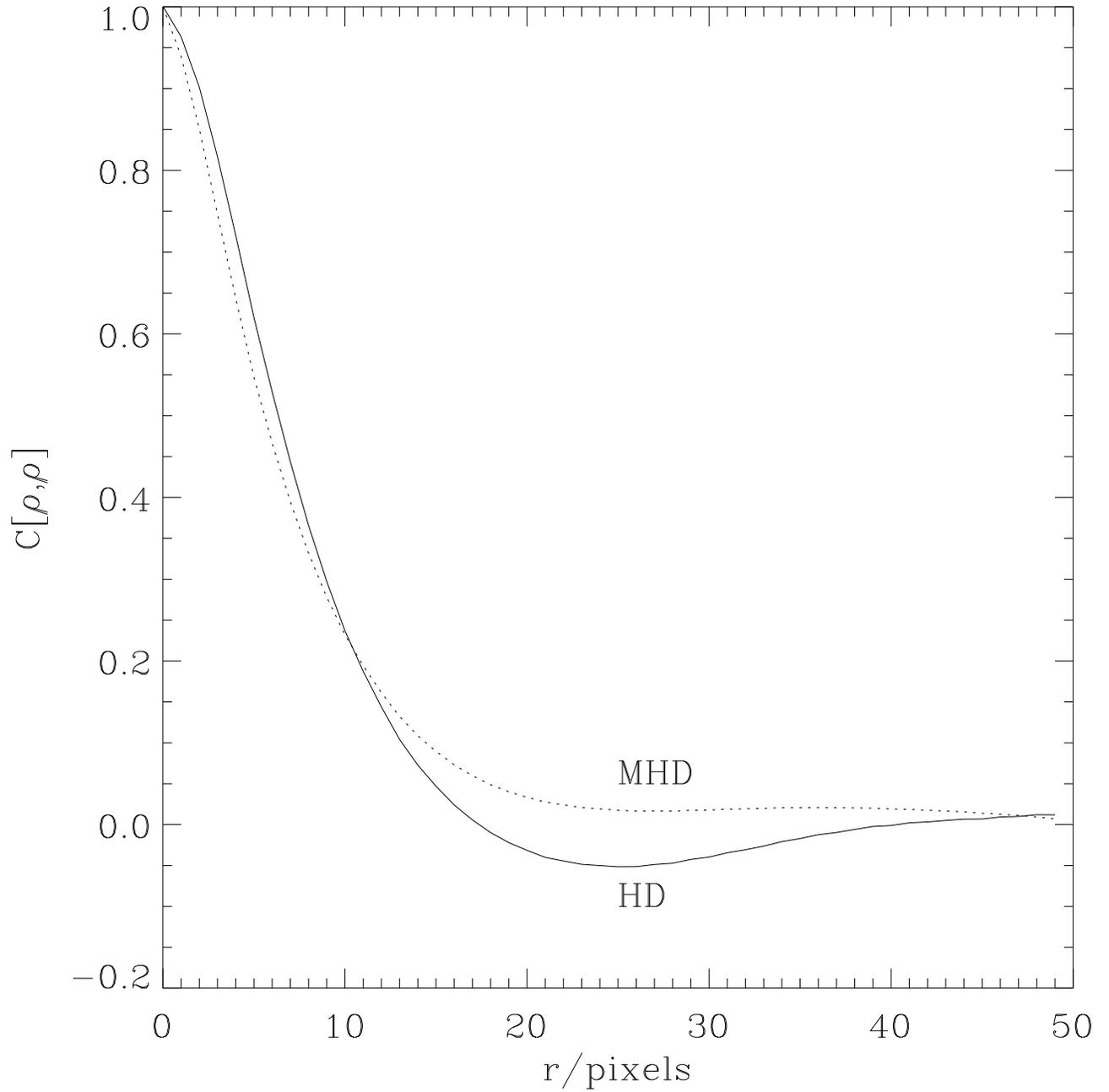}
\caption{Density autocorrelation function (ACF) for the non-magnetic ({\it 
solid line}) and magnetic ({\it dotted line}) simulations as a
function of separation (or ``lag'') $r$ in grid cells. The $r$ axis
extends to only  half the simulation size, because the periodic boundary
conditions imply that the curve is symmetric about this point.}
\label{fig:rho_corr}
\end{figure}

In order to test this hypothesis, we have computed the
ACF of the density field in the numerical
simulations, at time $t=2.8$ for the non-magnetic run, and at
$t=3.2$ for the magnetic run ($\sim 0.45$ and 0.51 large-scale
turbulent crossing times, respectively). These are shown in fig.\
\ref{fig:rho_corr} as a function of spatial
separation (``lag'') in grid cells. Note that we show lags only up to half 
the simulation size, since the periodic boundary conditions imply that
the ACF is symmetric about this value. It is seen that the
ACF has decreased to half its maximum (zero-lag) value at
separations of only about 7 cells, and to 10\% at 
lags of only $\sim 14$ cells. We can effectively consider the latter to be
a ``decorrelation length'' for the simulations. Note that the presence of
the magnetic field does not seem to have 
an important effect on the decorrelation length. For distances
significantly larger than this decorrelation length, the effects of density
autocorrelation should be negligible, and the CLT should be 
applicable (see \S \ref{sec:discussion} for a discussion of possible
caveats). We do not
choose the more familiar $1/e$ criterion for the decorrelation length
because of two reasons. First,
the $1/e$ criterion is only truly meaningful for exponential decay laws, but 
in general the ACF does not decay in this form, and in fact crosses
zero at a finite lag in the non-magnetic run. Second, we are
interested in lags at which the ACF has become effectively negligible
compared to its zero-lag value, so that events separated by this
length can be assumed to be independent, and a factor of 1/10 seems more
appropriate for this purpose than one of $1/e$. For these reasons, we
also have chosen to refer to this as a ``decorrelation'' length, rather
than a correlation one. But in any case, this
choice is essentially arbitrary. In what follows, we denote  the
ratio of the column (or cloud) length to the decorrelation length by $\eta$. 

Our simulations clearly do not have a large enough number of independent 
events along the LOS for the column density PDF to approach a Gaussian,
as $\eta \sim 7$ in the simulation box. 
We thus have chosen to study this problem using simple random
realizations of lognormal fields, sacrificing the realistic
hydrodynamic origin of actual density data in favor of the ability to
control $\eta$ precisely, and to
generate much longer LOSs than can be attained with even the
largest presently available
computational resources in numerical hydrodynamical simulations. This
approach has been used in the past for simulating turbulent velocity
fields without the numerical expense of actual hydrodynamical
simulations (e.g., Dubinsky, Narayan \& Phillips 1995; Klessen 2000;
Brunt \& Heyer 2000). The 
main feature that is lost by doing this is the spatial correlation
that is inherently present in actual mass density fields, due to the
continuum nature of real flows. In any case, random, spatially
uncorrelated fields should constitute a best-case scenario for
studying the convergence to a Gaussian PDF. The presence of
correlations of a certain size in grid cells should increase the required
path lengths for convergence by a factor equal to this size, making
convergence even slower. For the random lognormal realizations, one
decorrelation length can be thought of as a single cell, so that the
integration length along the line of sight $\Delta z$ equals $\eta$ for
the random data.

\begin{figure}
\plotone{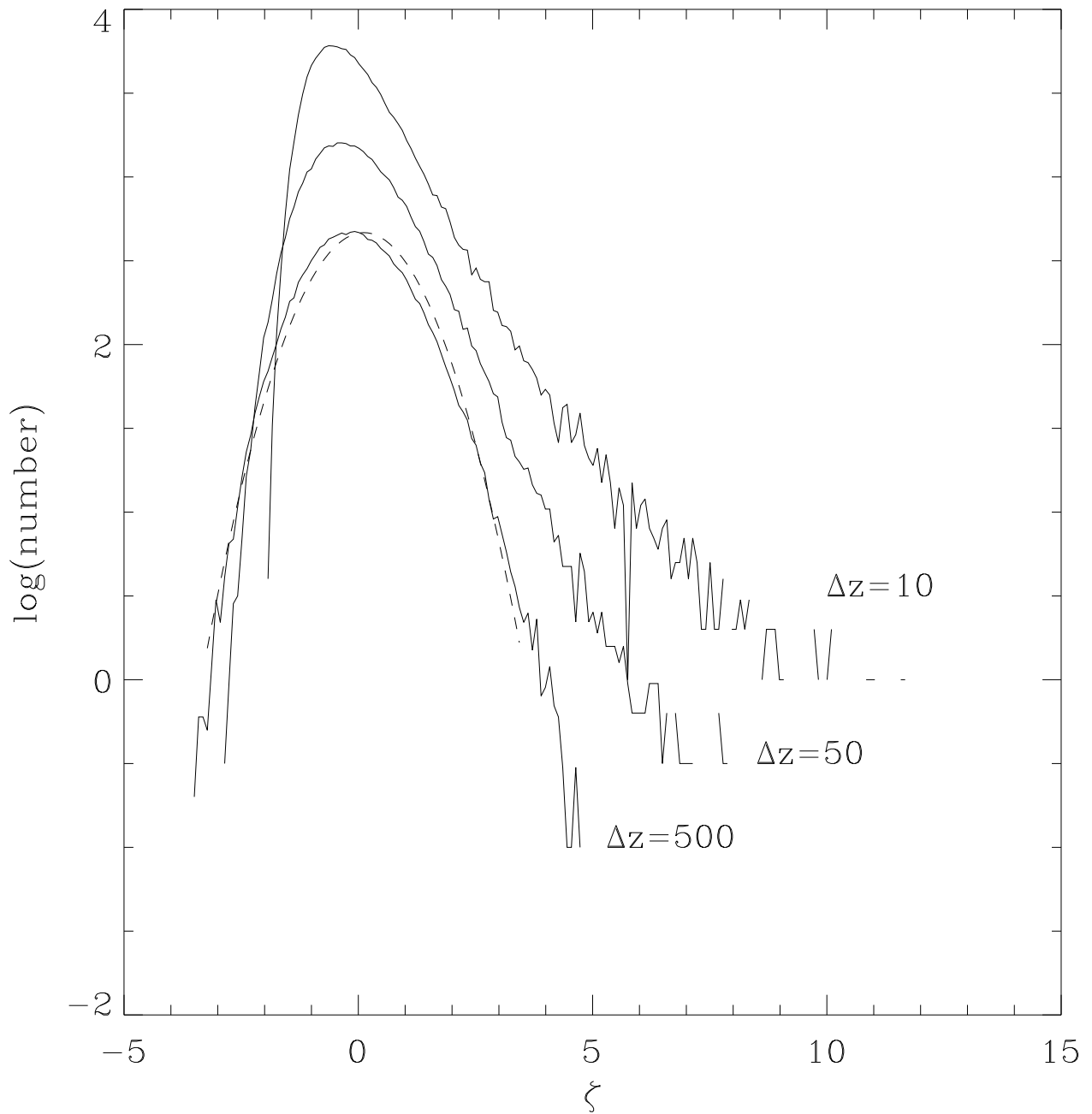}
\caption{Normalized column density ($\zeta$) PDFs of the random
lognormal density 
realizations, with $b=1$, and path (line-of-sight) lengths
$\Delta z= 10$, 50 and 500 grid cells, as indicated. The dashed line is a
Gaussian fit to the $\Delta z=500$ PDF over the $\zeta$-range
spanned by the dashed curve. Note the transition
from a nearly lognormal to a nearly Gaussian curve as $\Delta z$
increases. The PDF for
the intermediate case $\Delta z=50$ is nearly exponential.}
\label{fig:N_pdfs_b1}
\end{figure}

We study the convergence of the PDF to a Gaussian as a function of two 
parameters: the width of the underlying lognormal density PDF, given
by $b$, and $\Delta z$. Figure
\ref{fig:N_pdfs_b1} shows the
PDFs of $\zeta$ for three realizations with $\Delta z=$ 10, 50 and 500
grid cells. It can be seen that at $\Delta 
z=10$, the PDF of $\zeta$ appears to decay exponentially for $0 < \zeta< 4$,
but develops a concavity at larger $\zeta$. At $\Delta z=50$, the
high-$\zeta$ side of the PDF is almost a perfect exponential, but at
$\Delta z=500$ no exponential segment is left, and the curve begins to 
approach a Gaussian.

\begin{figure}
\plotone{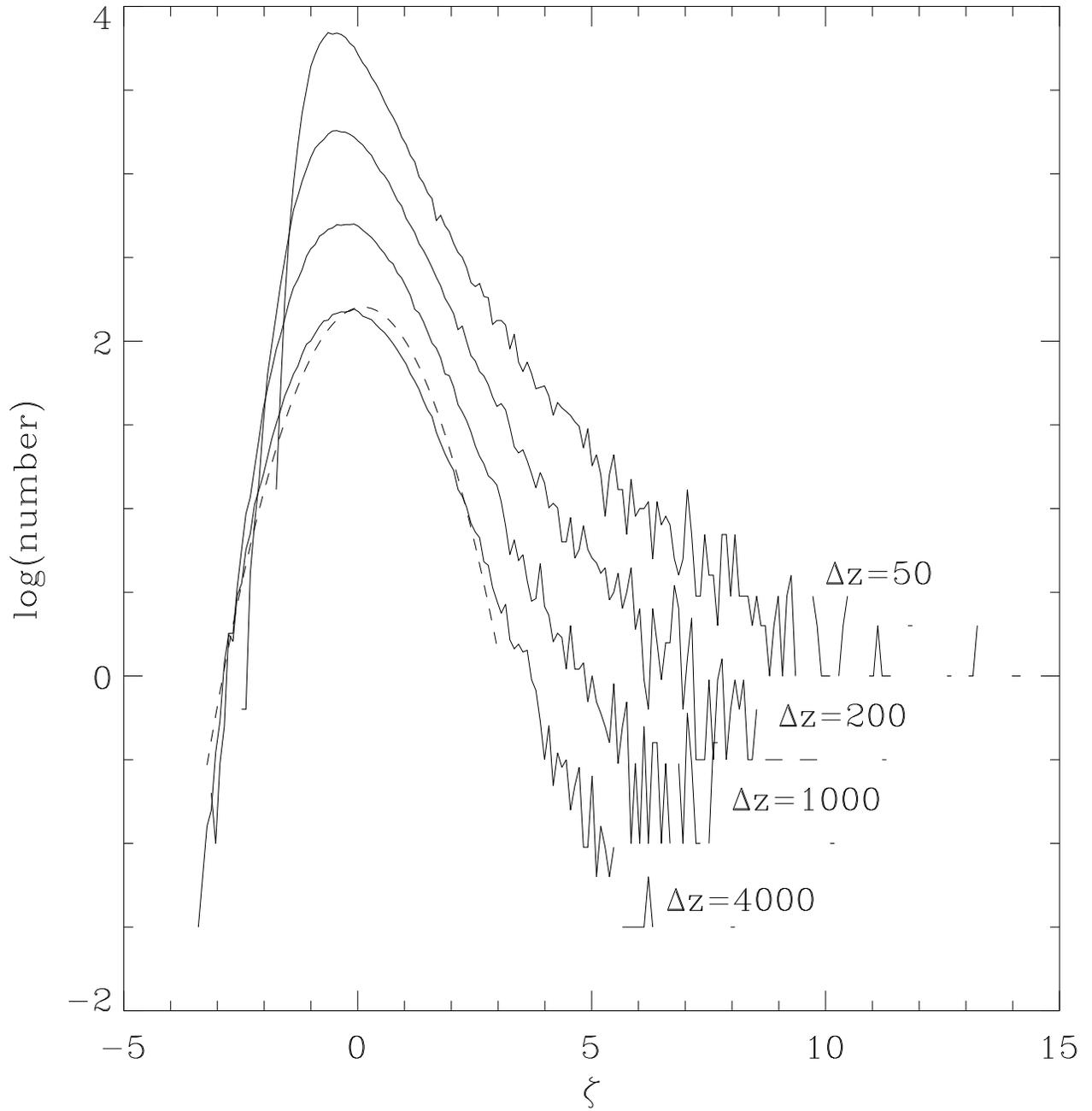}
\caption{Same as fig.\ \ref{fig:N_pdfs_b1} but for $b=1.5$ and path
lengths $\Delta z=50$, 200, 1000 and 4000 grid cells. Only at the latter
value does the nearly exponential behavior at moderately large $\zeta$
begin to disappear. The dashed line is a
Gaussian fit to the $\Delta z=4000$ PDF.}
\label{fig:N_pdfs_b1.5}
\end{figure}

Figure \ref{fig:N_pdfs_b1.5} shows a similar sequence as that in fig.\ 
\ref{fig:N_pdfs_b1}, but for $b=1.5$. In this figure we show
realizations with $\Delta z=$200, 500, 2000 and 4000. Again a
transition from concavity to convexity is seen to
occur at high $\zeta$ as $\Delta z$ increases, although in this case, even 
at $\Delta z=4000$ an excess is seen at the largest values of $\zeta$, so the
convergence is not yet complete at this path length for
$b=1.5$. Indeed, it is known that for very asymmetric 
distributions with important wing excesses, the convergence to a Gaussian is
fastest near the middle of the PDF, and slowest at the tails (Peebles
1987, sec.\ 4.7).
We conclude that, even for completely uncorrelated data, convergence to a
Gaussian occurs very slowly at the high-$\zeta$ tail if the
skewness, kurtosis and dynamic range of the underlying density data
are large. Also, we can expect that, as more LOSs are included in the 
column density PDF, the extreme-$\zeta$ tail will reach higher $\zeta$
values, and will thus require larger lengths $\Delta z$ to converge 

\section{Discussion} \label{sec:discussion}

\subsection{What is the value of $\eta$ in real molecular clouds?}
\label{sec:what_eta?} 

The convergence studied in the last section refers to completely
uncorrelated random data, so that the correlation length is effectively
one grid cell. As mentioned in \S \ref{sec:pdfs}, the
presence of a finite decorrelation length in the density
data should cause the convergence to be even slower with path length, as
sufficiently independent ``events'' are expected to be separated by lags of the
order of the decorrelation length. Note that, if the decorrelation length
is a sizeable fraction of the column (cloud) size, i.e., {\it if $\eta$ is
not very large, then full convergence to a Gaussian is not expected}.

Clearly, in the case of real molecular clouds, the concept of
``grid cell'' disappears, and the natural unit for measuring the path
length should be the decorrelation length itself. Thus, the ratio
$\eta$  serves as a measure of the path length. In our simulations,
$\eta \sim 7$, and, according to the results of \S \ref{sec:pdfs}, this
is too small a length to produce a full convergence to a Gaussian column
density PDF even at moderate underlying density contrasts. 
However, in real molecular clouds the actual value of $\eta$ is
essentially unknown, and convergence to a Gaussian column density PDF is
plausible, if $\eta$ is large enough. Thus, it becomes important to assess
this possibility. 

The value of $\eta$ must be related to some characteristic scale of the 
{\it density-fluctuation} power spectrum. If the spectrum has a
self-similar (power law) 
dependence on wavenumber over some range, then there are no
characteristic scales in this range\footnote{We thank Thierry Passot for 
pointing this out.}, and the only natural characteristic scales are
those where the power-law range ends at high and low wavenumbers
(analogous to the the ``inner'' and ``outer'' scales of the turbulent
energy spectrum). In order to investigate this dependence, we use a
spectrum-modifying algorithm introduced by Lazarian et al.\ (2001),
which allows us to modify the spectrum of any physical field without
modifying its spatial distribution. This amounts to only changing the
``contrast'' of the physical field (Armi \& Flament 1985). Since the
power spectrum only depends on the Fourier amplitudes of a field, but
not its Fourier phases, the 
modification is accomplished by Fourier-transforming the physical field, 
and then replacing the Fourier amplitudes by others that satisfy the
desired spectral shape, without altering the phases. We refer the reader 
to Lazarian et al.\ (2001) for details of the algorithm.

\begin{figure}
\plottwo{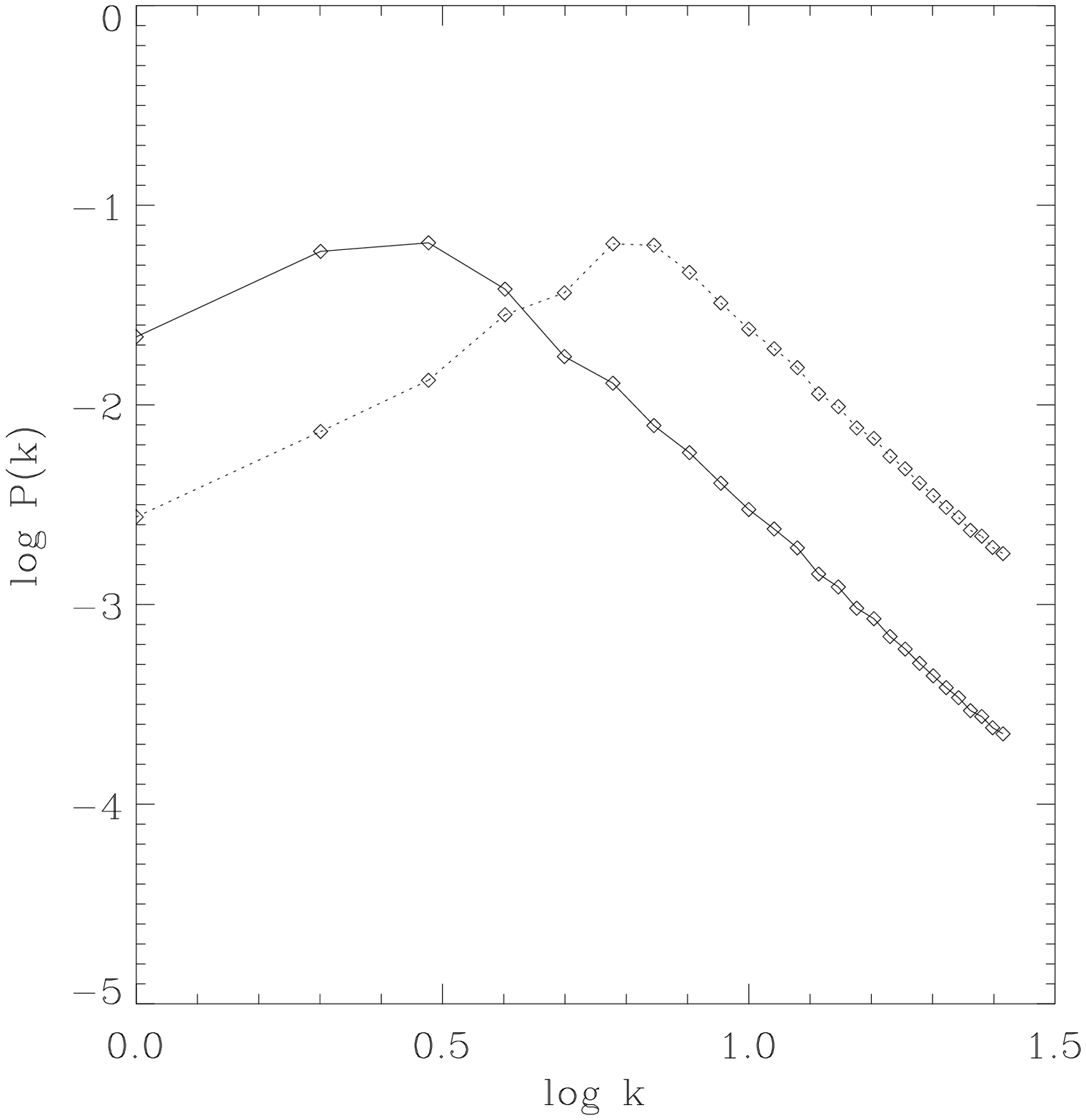}{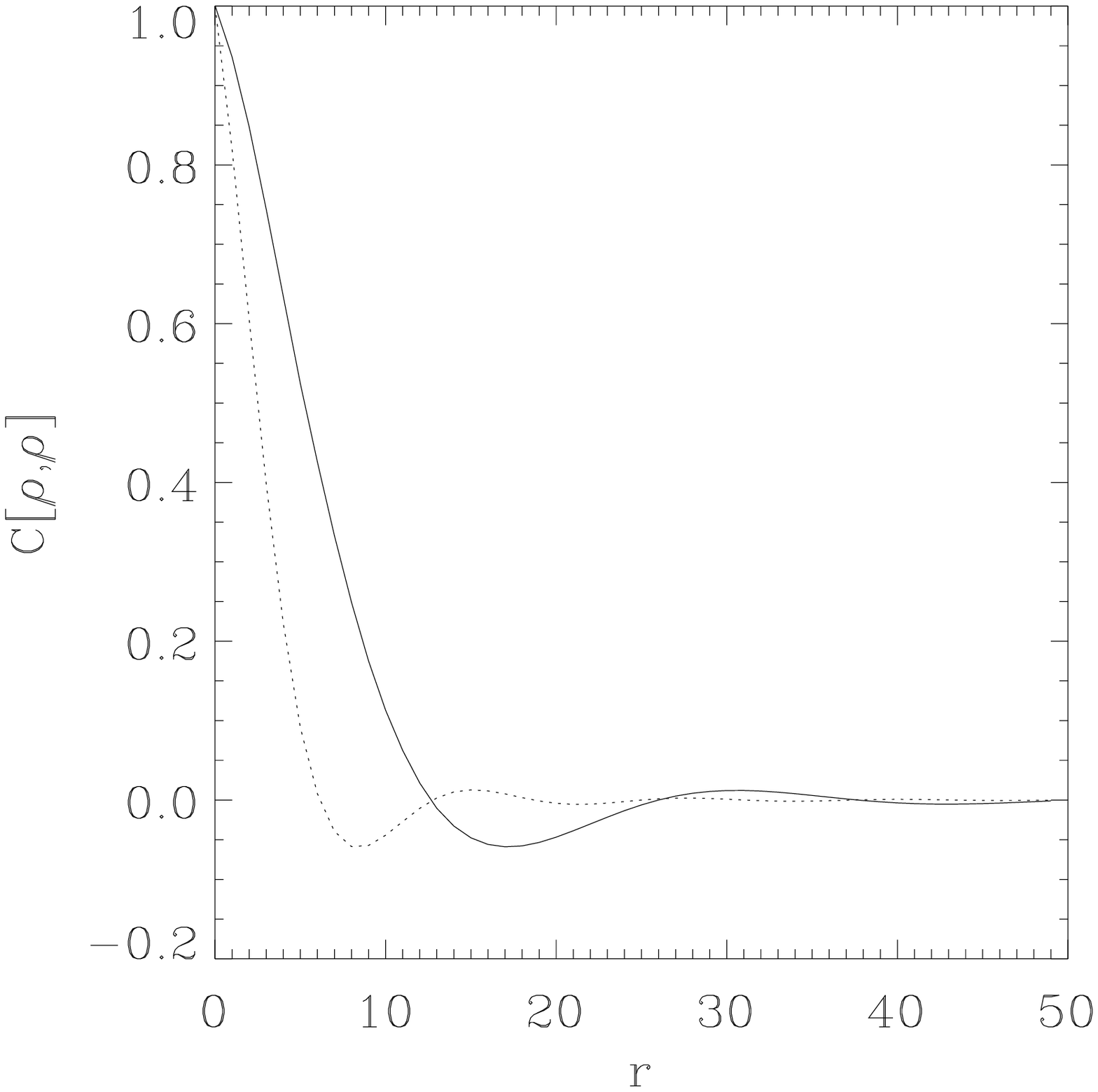}
\caption{{\it a) (left)} Two power spectra imposed on the density field
of the non-magnetic 
simulation for studying their effect on the decorrelation length. In
both cases the spectrum rises as $k^3$ until $k_{\rm p}$ and
then decreases as $k^{-3}$, where $k$ is the wavenumber in units of the
inverse box length. {\it Solid line:} $k_{\rm p}=3$; {\it dotted line:}
$k_{\rm p}=7$. {\it b) (right)} Resulting 
density autocorrelation functions for the power spectra shown in {\it
a)}. The line type matches that of the corresponding power spectrum.}
\label{fig:pow_acf}
\end{figure}

We apply the spectrum-modifying algorithm to the density field of the
non-magnetic simulation, and impose on it the two spectra shown in fig.\
\ref{fig:pow_acf}a. In both cases, the spectrum rises as $k^3$ for
$k \le k_{\rm p}$ and then decreases as $k^{-3}$ for $k > k_{\rm p}$,
where $k$ is the wavenumber in units of the inverse box length, and
$k_{\rm p}$ is a ``peak'' wavenumber. One case has $k_{\rm p}=3$ ({\it
solid line}) and the other has $k_{\rm p}=7$ ({\it dotted line}). 
These spectra produce the density ACFs shown in fig.\
\ref{fig:pow_acf}b. It is seen that the 10\% level of the ACF
occurs at a lag $r \sim 1/(2 k_{\rm p})$, suggesting that the
decorrelation length is related to the ``outer scale'' of the density
power spectrum.

What determines the shape of the density fluctuation power spectrum,
and, in consequence, of the density ACF in
highly compressible turbulence is, to our knowledge, an open problem. It 
should most likely be related to the energy spectrum and the forcing
(energy injection) 
spectrum\footnote{We thank E.\ Ostriker for noting this point.}, but the 
actual forms of all these spectra in molecular clouds are
unknown. For example, the temporally and spatially intermittent energy
injection in molecular clouds from embedded stellar sources and passing
shocks differs significantly from the standard random forcing scheme
used in most numerical simulations, which is applied everywhere in space 
and continuously in time
(see, e.g., Norman \& Ferrara 1996 and Avila-Reese \& \VS\ 2001 for related
discussions at larger scales in the ISM). In particular, if the energy
is injected at small scales, the energy spectrum in a cloud complex may
peak at scales quite smaller than the complex's size, and possibly drive 
the density field to a similar spectral shape. Moreover, note that,
if the energy spectrum is dominated by shocks, then its form ($k^{-2}$)
is of geometrical, rather than dynamical, origin (see, e.g., \VS\ et
al. 2000), and in this case the density power spectrum need not have the 
same outer scale as the the energy spectrum. Self-gravity may be an
important ingredient, too. In summary, the actual
shapes of all the relevant spectra in molecular clouds, and thus the
value of $\eta$, remain unknown, and deserve to be studied
systematically. 

Observationally, several workers have looked at correlation lengths
in molecular gas. In a pioneering study, Kleiner \& Dickman (1984)
investigated the ACF of column density in the Taurus region, and from
their plots one infers a correlation length of a few pc. This is not
too short a distance compared to the complex's size, but note, however, that
this correlation length refers to the projected intensity data rather
than to the underlying 3D density field. Most other
observational correlation 
studies have focused on the ACF of the line velocity centroid
distribution, and are not directly applicable to our puposes. In any case,
they have either reported correlation lengths of fractions of a parsec
(e.g. Scalo 1984; Kleiner \& Dickman 1985) or else find them difficult 
to determine unambiguously (e.g. Miesch \& Bally 1995).

In this respect, our results suggest that the column density
PDF provides us with a means of observationally measuring the ratio of
the cloud size to the decorrelation length $\eta$ when optically thin
transitions or extinction data are used: the observed column density
PDF should transit from a lognormal to an exponential and then on to a
Gaussian as $\eta$ increases. Unfortunately, we do not know the
path length a priori, but if it can be estimated by
some other means at least in some cases, then the decorrelation length,
and consequently the density power spectrum outer scale can be
derived. This suggests that it is necessary
to investigate numerically how the decorrelation length depends on
parameters of the flow such as the forcing parameters, self-gravity,
the energy and magnetic spectra, etc.

\subsection{The width of the column density PDF} \label{sec:PDF_width}

\begin{figure}
\plottwo{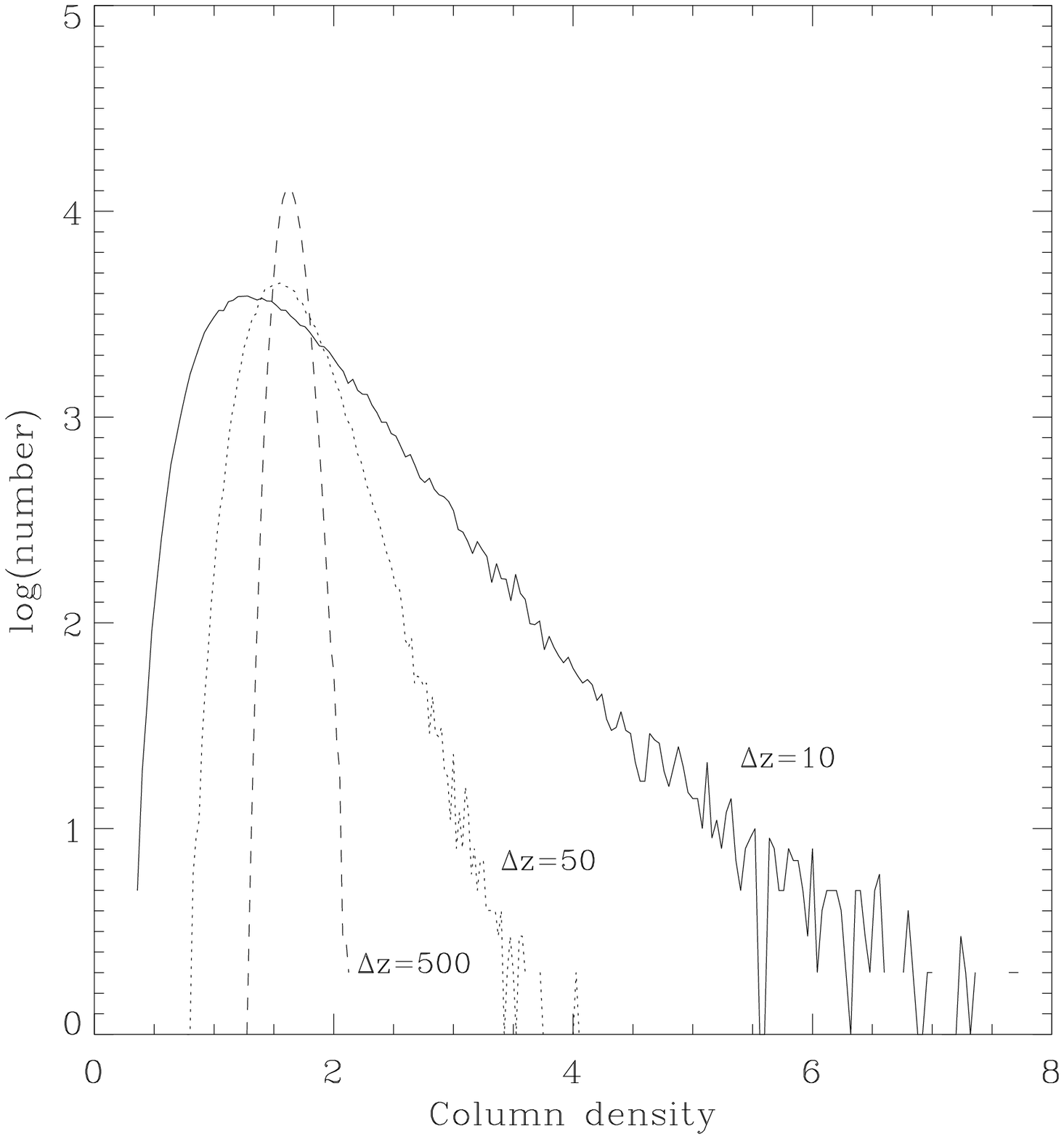}{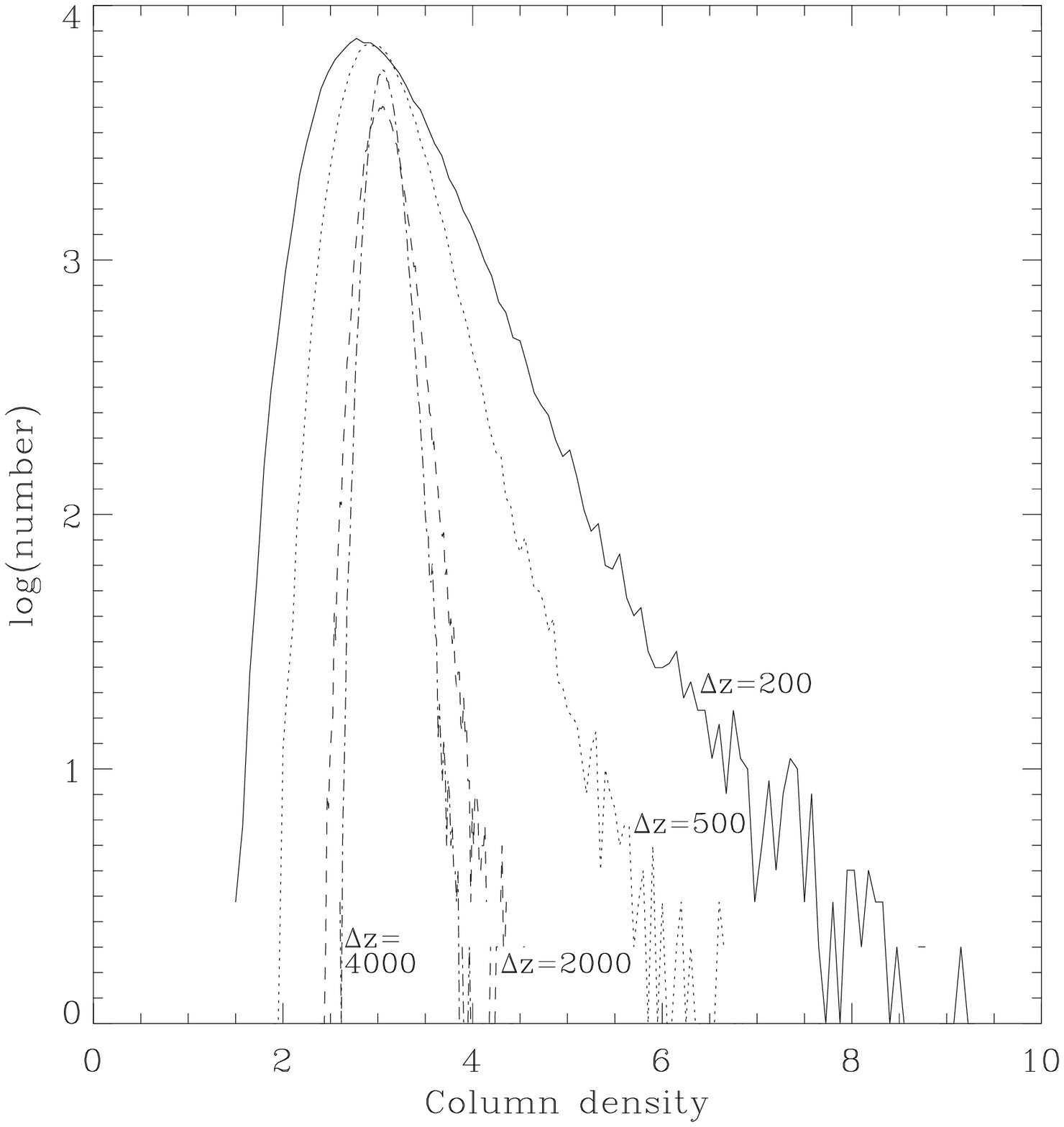}
\caption{{\it a) (Left)} PDFs of the {\it mean density} along every LOS
(i.e., un-normalized column density divided by path length) for the
random density realizations with $b=1$. For $\Delta z=10$ grid cells, the 
column density is seen to span a range of roughly a factor of 20, from 
0.4 to 8. For $\Delta z=500$, the range has been reduced to less than
a factor of 50\%. {\it b) (Right)} Same as in (a) but for $b=1.5$. In
this case, the column density range at $\Delta z = 200$ is a factor of
$\sim 6$.}
\label{fig:N_pdfs_nonstd}
\end{figure}

Another implication of the results from \S \ref{sec:pdfs} is that, at
large $\eta$, the column density
dynamic range becomes small. Figure \ref{fig:N_pdfs_nonstd} shows the
PDFs of the {\it mean density} (i.e., the {\it un-normalized} column density
divided by the path length) for all LOSs for the two sets of random density
fields. It is seen that, while the underlying
density PDFs discussed here have density contrasts of up to $10^6$,
the column density PDFs typically have dynamic ranges of at most a
factor of 20, and, for very large $\eta$, of only factors of a
few. This is actually a trivial
result, since in the limit $\eta \rightarrow \infty$, all LOSs
would give exactly the same column density (i.e., the sample mean
asymptotically approaches the distribution mean), and the column
density PDF would collapse to a Dirac delta function, independently of the 
dynamic range of the underlying density distribution. This suggests that,
if $\eta$ is large in
actual clouds, then nearly constant column densities are expected, but 
this tells little about the dynamic range of the actual
density field. In this case, Larson's (1981) density-size relation,
$\rho  \sim R^{-1}$, which implies constant column density,
could simply  be an observational averaging effect along the LOS (J.\
Scalo 2000, private communication). 
On the other hand, a relatively large observational column density
range would point towards relatively small values of
$\eta$. 

Observational studies of 
extinction (e.g., Lada et al.\ 1994; Kramer et al.\ 1998; Cambr\'esy
1999) typically report extinction (proportional to column density)
dynamic ranges of about a factor of 10. Comparing with the
mean-density PDFs of fig.\ \ref{fig:N_pdfs_nonstd}, these ranges are
consistent with $\eta \sim 10$ and $\eta \sim 100$ for
underlying density ranges of $10^4$ and $10^6$, respectively. For
comparison, the column densities reported by Padoan et 
al.\ (2000) from numerical simulations of MHD
turbulence at a resolution of $128^3$ with underlying density fields
with a dynamic range of $10^6$, span 3 orders of magnitude, suggesting
that in actual molecular clouds $\eta$ may be
significantly larger than in those
simulations. On the other hand, OSG01 have compared the 
column density cumulative distribution from their simulations to that
of visual extinction in cloud IC5146 (Lada, Alves \&
Lada 1999), finding
that the overall curve width in both data sets is comparable.
Unfortunately, in order to determine whether similar PDF widths imply
similar values of $\eta$, it is necessary to know whether the dynamic
ranges of the underlying density distributions are also
comparable. Moreover, due to the poorly-sampled nature of the observational
data, OSG01 had to present the distributions in
cumulative form, and in linear plots, rather than
semi-logarithmic. In this format, it would be
hard to distinguish, for example, between lognormal and exponential
PDFs that have similar values at moderate column densities. In
any case, the results are promising, and indicate that properties of
molecular cloud turbulence such as the decorrelation length may indeed
be determined from the column density PDF and the dynamic range of the
density field. 

Finally, note also that our results imply that there should
exist a relationship between the functional shape of the column
density PDF and its width, i.e., between its skewness and its
variance. We plan to quantify this relation in future work.

\subsection{The case of non-isothermal gas} \label{sec:non_iso}

In this paper we have restricted the analysis to lognormal
underlying density PDFs, in part for simplicity and in part in order
to relate our results on PDFs to those from recent numerical
simulations of compressible isothermal MHD turbulence (e.g., OSG01,
Padoan et al.\ 2000). Isothermal flows are normally 
considered as representative of the flow within molecular clouds.
However, it is possible that molecular is really only close to
being isothermal in the density range $10^3 \lesssim n/{\rm cm}^{-3}
\lesssim 10^4$ (see
the discussion by Scalo et al.\ 1998). Moreover, diffuse gas in the ISM,
either neutral or ionized, is in general non-isothermal, and in this
case, if the flow behaves approximately barotropically ($P \propto
\rho^\gamma$, $\gamma \ne 1$), a power-law range is expected to appear
in the PDF (Passot \& V\'azquez-Semadeni 1998). In this case the CLT
does not necessarily apply. Indeed, let us consider a power-law range of the
form $f(\rho)= C \rho^{-\alpha}$, where $C$ is a constant. If the
range extends to arbitrarily large and/or small values, the variance does
not exist, and therefore the CLT does not apply. If the power law is
truncated at low densities, and $\alpha>1$, then the column density
PDF becomes a gamma distribution (Adams \& Fatuzzo
1996). However, if the power-law range
has a finite extension, and beyond it the PDF drops rapidly, such as
the PDFs reported by Scalo et al.\ (1998) for non-isothermal 
numerical simulations of the ISM, and by Passot \& V\'azquez-Semadeni
(1998) for polytropic flows with $\gamma \ne 1$, then the
variance should still exist and the CLT should apply. We expect this to be
the case of observational PDFs of diffuse gas.

\subsection{Caveats} \label{sec:caveats}

Although the results of this paper are rather straightforward, a
number of possible complications should be kept in mind. First, it is
possible that the ACF fails to capture long-range correlations because 
the short-range ones may mask them, as small-scale structures are
generally much denser. So, even in cases where the 10\% level of the ACF 
is reached over lags much smaller than the cloud size, it is possible
that the flow is not sufficiently decorrelated for the CLT to apply (J.\ 
Scalo 2000, private communication).
Numerical simulations of turbulent flows with large
values of $\eta$ are necessary to test for this possibility. Second, in
cases where the Jeans length is close to the system size,
self-gravity may promote the formation of large-scale structures,
counteracting the possible action of small-scale energy injection
sources, and tending to reduce the value of $\eta$. In this case, column 
densities closer to lognormal shapes, and with rather large variances
might be expected. High-resolution numerical experiments with
self-gravity and realistic stellar-like forcing, even if just in 2D,
similar to 
those of Passot, \VS\ \& Pouquet (1995) or of \VS, Ballesteros-Paredes
\& Rodr\'\i guez (1999) but with cooling functions appropriate for molecular
clouds, may help resolve this issue.

Finally, we have suggested that a small column density dynamic range
should be taken as an indication of large values of
$\eta$. Unfortunately, small column density dynamic ranges 
may also arise from limitations in the sensitivity of the observations 
and saturation effects. Thus, the best suited observations for testing 
the above results are those in which these limitations are minimized.

\section{Summary} \label{sec:summary}

Our results can be summarized as follows:

\noindent
1. We have proposed that the relevant parameter determining the form of
the column density PDF in molecular clouds is the ratio $\eta$ of the
cloud size to the decorrelation length of the density field, with the
latter operationally defined in this paper as the lag at which the
density autocorrelation function (ACF) has decayed to its 10\%
level. Assuming that density ``events'' along the LOS are essentially
uncorrelated, large values of this ratio imply that the Central Limit
Theorem (CLT) can be applied to those events, and a Gaussian PDF should
be expected for large enough values of $\eta$. This parameter is
essentially the number of independent events (the ``sample size'') along 
the LOS, and the column density is equivalent to the ``sample mean''
along the LOS.

\noindent
2. We have measured $\eta$ in two 3D numerical simulations of isothermal 
turbulence forced at intermediate scales, one magnetic and one
non-magnetic. In both cases we find $\eta \sim7$, suggesting that at
least partial convergence to a Gaussian PDF should occur. The column
density PDFs for both runs are approximately exponential.

\noindent
3. Using simple random realizations of uncorrelated,
lognormally-distributed fields, we have shown that the PDF of the {\it
normalized} (i.e., with 
zero mean and unit variance) column density $\zeta$ indeed converges to a
Gaussian shape as $\eta$ increases, as dictated by the CLT,
albeit very slowly, due to the large dynamic range, skewness and
kurtosis of the density lognormal distribution. For cases in which the
underlying data have a dynamic range (``density contrast'') $\sim
10^4$, convergence to a Gaussian requires $\eta \sim$ several hundreds.
For density dynamic ranges $\sim 10^6$, the required sample size
is several thousand events. Additionally, the width (variance) of the column
density PDF also decreases as $\eta$ increases, as expected for the
distribution of the ``sample mean''. Specifically, for $\eta \sim 10$
and an underlying density dynamic range of $10^4$, the column density
dynamic range is $\sim 20$, and has decreased to a factor of a few for
$\eta \sim$ a few hundred.

\noindent
4. We have discussed the turbulent parameters that determine
$\eta$. Using a spectrum-modifying algorithm, we have shown that the
10\%-level decorrelation length appears to be given approximately by
$1/(2 k_{\rm p})$, where $k_{\rm p}$ is the wavenumber at which the
density fluctuation power spectrum peaks. Thus, the decorrelation length 
appears to be very close to the ``outer scale'' of the density power
spectrum. However, we believe that what determines the shape of the
density spectrum in molecular cloud turbulence is still an open problem
requiring much further work.

\noindent
5. We have suggested that the slow convergence of the column density PDF, 
which transits from lognormal (or a power-law, if the underlying gas 
behaves polytropically, but is not isothermal), to exponential and on to
nearly Gaussian
shapes as $\eta$ increases, can be used to observationally determine the 
latter in molecular clouds. This would provide a direct observational
diagnostic of this fundamental property of the turbulence in molecular
clouds. Additionally, since the variance of the column density PDF
decreases with increasing $\eta$, a functional relationship between the PDF's
variance and skewness is expected to exist. 

\noindent
6. The decrease of the PDF variance with increasing $\eta$ suggests
that, if $\eta$ turns out to be large in real molecular clouds, Larson's
(1981) density-size relation, which 
implies roughly constant column density, could be simply a result of this
averaging along the LOS (J.\ Scalo, 2000, private communication).
Conversely, wide, skewed PDFs may be an indication that the clouds are 
not very large compared to the turbulent density decorrelation length,
and Larson's relation might then be the result of limited observational
dynamic range (Kegel 1989; Scalo 1990; \VS\ et al.\ 1997).

\noindent
7. We also discussed briefly the case of power-law underlying density
PDFs, expected when the gas is not isothermal. In this case,
the CLT is only expected to apply if the power laws are truncated at
both low and high densities, although the convergence to a Gaussian
may be even slower if the power-law range is very extended, as power
laws have even higher tails than a lognormal distribution. 

\noindent
8. Finally, we have mentioned several possible caveats,
specifically: a) the possibility that the large-scale correlations are
masked in the density ACF because they involve lower-density
structures, b) the fact that self-gravity may possibly increase the
decorrelation length, and c) the fact 
that sensitivity and saturation problems with the observations
limiting their dynamic range may incorrectly be taken to mean large
values of $\eta$.

\acknowledgements

We thank Laurent Cambr\'esy for discussing his data with us, and Eve
Ostriker, Thierry Passot, Luis Rodr\'\i guez and John Scalo for sharp
comments and/or a careful reading of the manuscript. In particular, John
Scalo provided us with
important comments about statistical distributions, limitations
of the various statistical methods, and interesting
implications of this work. Remarks from Eve Ostriker helped us uncover a
misconception in an 
earlier version of this paper. The turbulence simulations were
performed on the Cray Y-MP 4/64 of DGSCA, UNAM. This work has made
extensive use of NASA's Astrophysics Data System Abstract
Service, and received partial funding from Conacyt grant 27752-E to E.\
V.-S.

\vfill
\eject

\begin{thebibliography}{99}
\bibitem[]{}Adams, F. C. \& Fatuzzo, M. 1996, ApJ 464, 256
\bibitem{} Armi, L. \& Flament, P. 1985, J. Geophys. Research,
90, no.\ C6, 11779
\bibitem[]{}Avila-Reese, V. \& \VS, E. 2001, ApJ 553, in press
(astro-ph/0101397)
\bibitem[]{}Brunt, C. \& Heyer, M. H. 2000, ApJ, in press (astro-ph/0011200)
\bibitem[]{}Cambr\'esy, L. 1999, A\&A 345, 965
\bibitem[]{}Dubinsky, J., Narayan, R. \& Phillips, T. G. 1995, ApJ 448, 226
\bibitem[]{}Gotoh, T. \& Kraichnan, R. H. 1993, Phys. Fluids A 5, 445
\bibitem[]{}Kleiner, S. C. \& Dickman, R. L. 1984, ApJ 286, 255
\bibitem[]{}Kleiner, S. C. \& Dickman, R. L. 1985, ApJ 295, 466
\bibitem[]{}Klessen, R. S. 2000, ApJ 535, 869
\bibitem[]{}Kramer, C., Alves, J., Lada, C., Lada, E., Sievers, A.,
Ungerechts, H. \& Walmsley, M. 1998, A\&A 329, L33
\bibitem[]{}Lada, C. J., Lada, E. A., Clemens, D. P. \& Bally,
J. 1994, ApJ 429, 694
\bibitem[]{}Lada, C.J., Alves, J. \& Lada, E. A. 1999, ApJ, 512, 250
\bibitem[]{}Larson, R. B. 1981, MNRAS, 194, 809
\bibitem[]{}Lazarian, A., Pogosyan, D., \VS, E. \& Pichardo, B. 2001,
ApJ, 554, in press (astro-ph/0102380)
\bibitem[]{}Miesch, M. S. \& Bally, J. 1995, ApJ 429, 645
\bibitem[]{}Nordlund, \AA \& Padoan, P. 1999, in Interstellar
Turbulence, eds. J. Franco and A. Carrami\~nana (Cambridge: Cambridge
University Press), p. 218
\bibitem[]{}Norman, C. A. \& Ferrara, A. 1996, ApJ 467, 280
\bibitem[]{}Ostriker, E. C., Gammie, C. F. \& Stone, J. M. 1999, ApJ 513, 299
\bibitem[]{}Ostriker, E. C., Stone, J. M. \& Gammie, C. F. 2001, ApJ,
546, 980 (OSG01)
\bibitem[]{}Padoan, P., Nordlund, \AA, \& Jones, B. J. T. 1997, MNRAS
288, 145
\bibitem[]{}Padoan, P., Juvela, M., Bally, J. \& Nordlund, \AA 2000,
ApJ 529, 259
\bibitem[]{}Passot, T., V\'azquez-Semadeni, E. \& Pouquet, A. 1995,
ApJ 455, 702
\bibitem[]{}Passot, T. \& V\'azquez-Semadeni, E. 1998, Phys. Rev. E
58, no. 4, 4501
\bibitem[]{}Peebles, P. Z., Jr. 1987, Probability, Random Variables, and Random
Signal Principles, 2nd ed. (New York: McGraw-Hill)
\bibitem[]{}Porter, D. H., Pouquet, A. \& Woodward, P. R. 1991, in
Large-Scale Structures in Hydrodynamics and Theoretical Physics,
eds. J. D. Fournier and P. L. Sulem (Berlin: Springer -Verlag), p. 105
\bibitem[]{}Scalo, J., V\'azquez-Semadeni, E., Chappell, D.\ \&
Passot, T., 1998, ApJ 504, 835
\bibitem[]{}Smith, M. D., Mac Low, M.-M. \& Zuev, J. M. 2000, A\&A 356, 287
\bibitem[]{}Smith, M. D., Mac Low, M.-M. \& Heitsch, F. 2000, A\&A 362, 333
\bibitem[]{}Stone, J. M., Ostriker, E. C. \& Gammie, C. F. 1998, ApJ 508, L99
\bibitem[]{}V\'azquez-Semadeni, E. 1994, ApJ 423, 681
\bibitem[]{}V\'azquez-Semadeni, E., Passot, T. \& Pouquet, A. 1996,
ApJ 473, 881
\bibitem[]{}\VS, E., Ballesteros-Paredes, J. \& Rodr\'\i guez,
L.F. 1997, ApJ 474, 292
\bibitem[]{}V\'azquez-Semadeni, E., Ostriker, E. C., Passot, T., Gammie, C.
\& Stone, J., 2000,  in Protostars \& Planets IV, eds. V. Mannings,
A. Boss \& S. Russell (Tucson: Univ.\ of Arizona Press), p. 3

\end{thebibliography}
\end{document}